
\magnification=1200
\def\C#1#2#3{C^{#1_0}_{#2_0 #3_0}}
\def\Z#1#2#3#4{Z^{#1_#3}_{#2_#4}}

\def\G#1{G_{#1_0}}

\def\e#1#2{\eta^{#1_{#2}}}
\def\P#1#2{{\cal P}_{#1_{#2}}}

\def\l#1#2#3{\lambda_{#1}^{{#2}\ {#3_#2}}}
\def\cb#1#2#3{\bar C_{{#1}\ {#3_#2}}^{#2}}
\def\b#1#2#3{b_{{#2}\ {#3_#2}}^{#1}}
\def\r#1#2#3{\rho_{#2}^{{#1}\ {#3_#2}}}
\def\A#1#2#3{A_{#1_#3}^{#2_#3}}
\rightline{ULB-TH-94/08}
\vskip2.5truecm
\centerline{\bf PATH INTEGRAL AND SOLUTIONS OF THE CONSTRAINT EQUATIONS:}
\centerline{\bf THE CASE OF REDUCIBLE GAUGE THEORIES.}
\vskip6pc
\centerline{Rafael Ferraro$^{(a,b)}$, Marc Henneaux$^{(c,d)}$ and
Marcel  Puchin$^{(e)}$}
\vskip5pc
\centerline{$^{(a)}${\it Instituto de Astronom\'{\i}a y
F\'{\i}sica del Espacio,}}
\centerline{\it C.C.67, Sucursal 28, 1428 Buenos Aires, Argentina,}
\vskip1pc
\centerline{$^{(b)}${\it Facultad de Ciencias Exactas y Naturales,
Departamento de
F\'{\i}sica,}}
\centerline{\it Universidad de Buenos Aires, Ciudad Universitaria, Pab.I, 1428
Buenos Aires, Argentina.}
\vskip1pc
\centerline{$^{(c)}${\it Facult\'e des Sciences, Universit\'e Libre de
Bruxelles, Campus Plaine,}}
\centerline{\it C.P.231, B-1050 Bruxelles, Belgium.}
\vskip1pc
\centerline{$^{(d)}${\it Centro de Estudios Cient\'{\i}ficos
de Santiago, Casilla 16443, Santiago 9, Chile.}}
\vskip1pc
\vskip1pc
\centerline{$^{(e)}${\it Universitatea din  Craiova,  Str.A.I.Cuza  13,
Facultatea  de  Fizica,  1100  Craiova, Romania.}}

\vskip6pc
\centerline{\bf ABSTRACT}
\vskip2pc
It is shown that the BRST path integral for reducible gauge
theories, with appropriate boundary conditions on the ghosts, is a
solution of the constraint equations.  This is done by relating the
BRST path integral to the kernel of the evolution operator projected
on the physical subspace.

\vfill

\eject

In the Dirac quantization of gauge systems, the physical states are
annihilated by the constraint operators
$$G_a (\hat q^i, \hat p_i)\ |\psi>\ =\ 0.\eqno(1)$$
In the
coordinate representation, the constraints,
which we shall assume for definiteness to be of even Grassmann parity, read
$$G_a (q^i, {\hbar\over i} {\partial\over\partial q^i})\ \psi(q^i)\ =\ 0.
\eqno(2)$$
The equations (1-2) are Gauss law constraints in Yang-Mills theory; while
in Einstein gravity, they are known as the Wheeler-De Witt equations.
\bigskip
The equations (1-2) are hard to solve in practice for
realistic theories. For this reason, it has been tried by several authors to
build solutions of (1-2) through path integral methods. This approach has
been initiated by Hartle and Hawking [1] in the case of quantum gravity, and
has been systematically studied in Ref.[2-4]. In particular it has been shown
on general grounds that the path integral with appropriate non minimal sector
and appropriate boundary conditions for the ghosts and the antighosts is
a solution of the constraint equations in the case of irreducible
constraints$^{\not= 1}$.
The purpose of this letter is to extend the results of [2-4] to the case of
reducible constraints. As it is well known, reducible constraints should also
be imposed as operator equations on the physical states in the Dirac
quantization
method, yielding as in (2)
$$\G{a} (q^i, {\hbar\over i} {\partial\over\partial q^i})\ \psi(q^i)\ =\ 0,
\eqno(3)$$
where we have replaced the index $a$ by $a_0$ since there will be further
indices$^{\not= 2}$.
Because of the reducibility of the constraints, however,
the ghost spectrum is much bigger and, in order to produce solutions
of (3) by path integral methods, one must specify the boundary conditions
that the new variables should fulfill. This is done in this letter. We shall
follow the method of Ref.[4], where the path integral is related to a
definite
operator expression, namely the ``projected kernel" of the evolution operator.
It should be stressed that the problem of
the boundary conditions on the ghosts and their
conjugate variables is non trivial.  Indeed,
different boundary conditions generically lead to a path integral
expression which may be of interest, but which would not solve the
constraint equations.
\bigskip
Our starting point is the concept of ``projected kernel" of a gauge invariant
operator $A_0 (\hat q, \hat p)$,$\ $ $[A_0, \G{a} ]\approx 0$ (Ref.[4],
Chap.13).
Let $|\psi_\alpha>$ be a basis of solutions of the constraint equations (3).
The scalar product $<\psi_\alpha|\psi_\beta>$ is generally infinite because it
envolves an integration over the gauge orbits. The physical scalar
product is obtained by inserting a gauge condition. We denote it by
$(\psi_\alpha|\psi_\beta)$ and we assume that the basis
$|\psi_\alpha>$ is orthonormal, $(\psi_\alpha|\psi_\beta) =
\delta_{\alpha\beta}$.

Any gauge invariant operator $A_0$ maps the
physical subspace on the physical subspace. Its physical matrix elements are
$(\psi_\alpha|A_0|\psi_\beta)$. The projected kernel of the operator
$A_0$ in the cordinate representation is defined by
$$A^P_0(q',q)\ =\ \sum_{\alpha,\beta}
<q'|\psi_\alpha> (\psi_\alpha|\hat A_0|\psi_\beta)
<\psi_\beta|q>,\eqno(4)$$
and is manifestly a solution of the constraint equations at $q'$ and
of the complex conjugate at $q$.
The projected kernel contains all the information about the physical matrix
elements of $A_0$ and fulfills the folding relation
$$(A_0\ B_0)^P\ (q',q)\ =\int dq' A^P_0(q'',q) \ \mu\left(q', {\hbar\over i}
{\partial\over\partial q'}\right)\ B^P_0(q',q),\eqno(5)$$
where $\mu$ is the insertion needed to regularize the scalar product,
$(\psi_\alpha|\psi_\beta) = <\psi_\alpha|\mu|\psi_\beta>$.
\bigskip
The projected kernel of $A_0$ may be related to BRST quantization.
To see this, it is necessary to recall some elements of BRST theory.
As standard we assume that it has been possible to order the ``minimal" BRST
charge $\Omega^{min}$ in such a way that
$$\left(\Omega^{min}\right)^2 \ = \ 0,\eqno(6)$$
and
$$\left(\Omega^{min}\right)^\dagger\ =\ \Omega^{min}.\eqno(7)$$
The operator $\Omega^{min}$ fulfilling (6-7) can be written in $\eta -
{\cal P}$ order (all ghosts $\e{a}{s}$ to the left of their conjugate
operators
$\P{a}{k}$) by using repeatedly the ghost (anti)commutation relations, to
yield
$$\Omega = \e{a}{0} \G{a} + \sum^{L-1}_{k=0}\e{a}{{k+1}} \Z{a}{a}{k}{{k+1}}
\P{a}{k} + ``more",\eqno(8)$$
where $``more"$ contains at least one $\eta$ and two ${\cal P}$'s, or two
$\eta$'s and one ${\cal P}$. Here, the $\e{a}{0}$ are the ghosts of the
zeroth generation associated with the constraints, the $\e{a}{1}$ are the
ghosts of ghosts of the first generation associated with the reducibility
identity $\Z{a}{a}{0}{1}\G{a}=0$, the $\e{a}{2}$ are the ghosts of ghosts
of the second generation associated with the reducibility identity
$\Z{a}{a}{1}{2}\Z{a}{a}{0}{1}\approx 0$, etc. The $\P{a}{k}$ are their
conjugate momenta.
\bigskip
As in [4], Section 14.5, we {\it define} the constraint operators of the Dirac
approach to be the coefficients of the ghost operators in (8) (see also
[5, 6]). This is not
the only possible definition but it has the following nice properties
\medskip
\noindent (i) In the limit $\hbar\rightarrow 0$, $\G{a}$ goes over into the
original classical constraints.
\medskip
\noindent (ii)
$[\G{a},\G{b}]=\C{c}{a}{b}\G{c}$ (in that order).
\medskip
\noindent (iii)
$\Z{a}{a}{0}{1}\G{a}=0$ (in that order).

\bigskip
\noindent
Because of (ii) and (iii), the equations $\G{a} |\psi> = 0$ are
quantum-mechanically reducible and fulfill the necessary integrability
condition for having solutions (``no anomaly").
\bigskip
Given a BRST invariant operator $A$ of ghost number zero,
$$\left[A,\Omega\right]\ =\ 0,\ \ \ \ \ \ \ \ \ \ \ A^\dagger = A,\eqno(9)$$
we define the gauge invariant operator $A_0(q,p)$ of which $A$ is the BRST
invariant extension, as the ghost-independent part of $A$ in its $\eta$-${
\cal P}$
ordering,
$$A\ = \ A_0 \ + \ \e{a}{0}\ \A{a}{b}{0}\ \P{b}{0}\ +\ ``\eta - {\cal P}".
\eqno(10)$$
It then follows from (9) that $[A_0,\G{a}] = \A{a}{b}{0} \G{b}$ (in that
order).
\bigskip
Under the unitary transformation generated by
$$C\ =\ {1\over 2} \left(\e{a}{0}\ \epsilon_{a_0}^{b_0}\ \P{b}{0}\ -
\P{b}{0}\ \epsilon_{a_0}^{b_0}\ \e{a}{0}\right) \
+ {1\over 2} \left(\e{a}{1}\ \epsilon_{a_1}^{b_1}\ \P{b}{1}\ +
\P{b}{1}\ \epsilon_{a_1}^{b_1}\ \e{a}{1}\right) \ +\ ...,\eqno(11)$$
the constraints and reducibility operators transform as
$$\G{a}\longrightarrow \bar\G{a}\ =\ \G{a} + \epsilon_{a_0}^{b_0}\ \G{b} +
{1\over 2} [\G{a},\epsilon_{b_0}^{b_0}] -
{1\over 2} [\G{a},\epsilon_{b_1}^{b_1}] + ...\eqno(12)$$

$$\Z{a}{a}{0}{1}\longrightarrow \bar\Z{a}{a}{0}{1}\ =\ \Z{a}{a}{0}{1} -
\Z{b}{a}{0}{1} \epsilon_{b_0}^{a_0}\  +
\epsilon_{a_1}^{b_1}\ \Z{a}{b}{0}{1}  +
{1\over 2} [\Z{a}{a}{0}{1},\epsilon_{b_0}^{b_0}] -
{1\over 2} [\Z{a}{a}{0}{1},\epsilon_{b_1}^{b_1}] + ...\eqno(13)$$
The physical states $|\psi>$ of the Dirac approach, solutions of (3),
transform thus as
$$|\psi>\longrightarrow |\bar\psi>\ =\ |\psi> -
{1\over 2} \epsilon_{a_0}^{a_0}\ |\psi> +
{1\over 2} \epsilon_{a_1}^{a_1}\ |\psi> - ...\eqno(14)$$
This leaves the physical scalar product unchanged (although $<\psi|\chi>$
may be modified), because the regulator $\mu$ in $(\psi|\chi)$ $=$
$<\psi|\mu|\chi>$ transforms as $\mu\rightarrow\bar\mu$ $=$ $\mu + {1\over 2}
(\epsilon\mu + \mu\epsilon)$ with $\epsilon = \epsilon_{a_0}^{a_0} -
\epsilon_{a_1}^{a_1} + ...$ [4]. In Ref.[7], it has been shown that the above
definition of quantum constraints and transformation laws are quite natural
for constraints linear in the momenta, and amount to regarding the wave
functions as densities of weight $1/2$ under changes of coordinates in $(q,
\eta)$ space.
\vskip2pc
The physical states of the BRST method are the states annihilated by the BRST
charge. It is easy to verify that the solutions of the constraints (3),
tensored by the ghost state annihilated by all the ghost momenta is a
solution of $\Omega^{min}|\Psi>=0$
$$|\Psi_\alpha>\ =\ |\psi_\alpha>\otimes |\chi>,\ \ \ \ \ \ \
\G{a} |\psi_\alpha>\ =\ 0, \ \ \ \ \ \ \ \ \P{a}{k} |\chi>\ =\ 0\ \ \ \ \
\Rightarrow \ \ \ \ \ \ \Omega^{min}|\Psi_\alpha>\ =\ 0.\eqno(15)$$
Furthermore, these states cannot be BRST-exact since they do not involve the
ghosts. For each solution of (3) there is thus
one and only one equivalence class
of BRST physical states.

The states (15) are convenient for making the comparison between the Dirac
and BRST quantization methods. However, it is as difficult to construct them
as
it is difficult to find the physical states of the Dirac method, since in both
cases one needs to solve explicitly
the constraint equations (3). For this reason it is
useful to replace the states (15) by states that are dual to them, namely
the states annihilated by the ghosts themselves,
$$\e{a}{k}\ |\Psi>\ =\ 0\ \ \ \ \ \ \ \ \ \ \Rightarrow \ \ \ \ \ \ \ \ \
\Omega^{min}|\Psi>=0.\eqno(16)$$
These states are solutions of the BRST physical state condition no matter
what their $q$-dependence is, since each term in the BRST charge
$\Omega^{min}$ contains one more $\eta$
than there are ${\cal P}$'s and therefore annihilates
$|\Psi>$. They are in general not independent in cohomology, however, since
some
of them are BRST exact. The cohomological classes that they define are
actually
dual to the cohomological classes defined by (15) (Ref.[4],\S 14.5.3;
see also [8,J9]). So,
one may say that while the constraints $\G{a} |\psi>=0$ are directly enforced
for the states (15), they are indirectly enforced for the states (16),
through the passage  to the BRST cohomology.

The states (16) do not have, in general, ghost number zero, ${\cal G}|\Psi>
\neq 0$. This leads to ill-defined scalar products.
For this reason, it is convenient to extend further the ghost spectrum, by
introducing new variables, and to tensor the states (16) with appropriate
states of the new sector, in such a way that the product states have total
ghost number zero [4, 8, 9]. The new variables
are called ''non minimal" variables.

In the reducible case, the non minimal variables are given by blocks
$\big[\left(\l{k}{s}{a}, \b{k}{s}{a}\right)$,
$\left(\cb{k}{s}{a}, \r{k}{s}{a}\right)\big]$, $0\le k \le L$,
$k\le s \le L$, where $\l{0}{0}{a}\equiv \lambda^{a_0}$ are the Lagrange
multipliers for the constraints and
$$[\b{{k'}}{{s'}}{b},\l{k}{s}{a}] = - \delta ^{k'}_k\
\delta ^s_{s'}\ \delta ^{a_s}_{b_{s'}}\ =\ [\r{{k'}}{s}{a},
\cb{k}{{s'}}{{b}}],\eqno(17)$$
$$\epsilon (\b{k}{s}{a})\ = \ \epsilon (\l{k}{s}{a})\ = \ s\ -\ k\ \ \ \
(mod\ \ 2),\eqno(18)$$
$$\epsilon (\r{k}{s}{a})\ = \ \epsilon (\cb{k}{s}{a})\ = \ s\ -\ k\ +\ 1\ \
\ \
(mod\ \ 2),\eqno(19)$$
$$gh(\b{k}{s}{a})\ =\ k\ -\ s\ =\ - gh(\l{k}{s}{a}),\eqno(20)$$
$$gh(\r{k}{s}{a})\ =\ -\ k\ +\ s\ +\ 1\ =\ - gh(\cb{k}{s}{a}),\eqno(21)$$
(see Ref.[10], and [4] Section 19.2.4).

The BRST generator is $\Omega = \Omega^{min}+\Omega^{non\ min}$ where
$$\Omega^{non\ min}\ =\ \sum_{k=0}^L\ \sum_{s=k}^L\ \b{k}{s}{a}\
\r{k}{s}{a}.\eqno(22)$$
The constant in the ghost number operator  ${\cal G} = {\cal G}^{min}+
{\cal G}^{non\ min}$ is adjusted so that ${\cal G}$ is antihermitian.
With this choice, ${\cal G}$ is a sum over all conjugate pairs of
terms of the form $i g \eta {\cal P} + (-)^{\epsilon_\eta} (g/2)$, where
$\epsilon_\eta$ is the parity of the conjugate pair $(\eta, {\cal P})$
in question.
\bigskip
We are now in the position to state the main results of our letter. There are:
\vskip1cm
\noindent{\bf Theorem 1.} The states $|q,\, ghosts>$ defined by
$$\eqalign{|q,\,ghosts>\ \equiv \ |q^i>\ \prod_{s=0}^L\ |\e{a}{s}=0>\
&\prod_{k odd}\ \prod_{s=k}^L\ |\l{k}{s}{a}=0,\r{k}{s}{a}=0>\cr
&\prod_{k even}\ \prod_{s=k}^L\ |\b{k}{s}{a}=0,\cb{k}{s}{a}=0>,\cr\cr}
\eqno(23)$$
are
\noindent (i) BRST invariant,
$$\Omega \; |q,\,ghosts>\ =\ 0,\eqno(24)$$
and
\noindent (ii) of ghost number zero,
$${\cal G} \; |q,\,ghosts>\ =\ 0.\eqno(25)$$
\bigskip
\noindent{\bf Theorem 2.} The matrix elements $<q',\, ghosts|A|q,\, ghosts>$
of the BRST invariant operator $A$ between the states of Theorem 1 are,
up to inessential numerical factors that we shall not write, just
equal the projected kernel (4) of the operator $A_0$ of which $A$ is the
BRST invariant extension,
$$<q',\, ghosts|A|q,\, ghosts>\ \sim \  A_0^P(q',q).\eqno(26)$$
\bigskip
In particular, if one takes for $A$ the evolution operator $U(t'-t)$ $=$
$\exp{-i H (t'-t)}$, where $H$ is the BRST
invariant extension of the Hamiltonian,
one gets, using the fact that the projected kernel is annihilated by the
constraints:
\bigskip
\noindent{\bf Theorem 3.} The path integral
$$\eqalign{U^P_0(q',t';q,t)\ =\ \int\ {\cal D}q^i\ {\cal D}p_i\ {\cal D}\e{a}
{s}\
&{\cal D}\P{a}{s}\ {\cal D}(non\ \ minimal\ \ sector)\cr & \exp{i\int_t^{t'}
dt
(p\dot q+\P{a}{s}\dot{\e{a}{s}}+.......-H)},\cr\cr} \eqno(27)$$
with boundary conditions
$$q^i(t')\ =\ q'^i,\ \ \ \ \ \ \ \ \ \ \ \ q^i(t)\ =\ q^i,\eqno(28)$$
$$\e{a}{s}(t')\ =\ 0,\ \ \ \ \ \ \ \ \ \ \ \ \e{a}{s}(t)\ =\ 0,\eqno(29)$$
$$\l{k}{s}{a}(t')\ =\ 0,\ \ \ \ \ \ \ \ \ \ \ \ \l{k}{s}{a}(t)\ =\ 0
\ \ \ \ \ \ \ \ \ \ k\ \ odd,\eqno(30)$$
$$\b{k}{s}{a}(t')\ =\ 0,\ \ \ \ \ \ \ \ \ \ \ \ \b{k}{s}{a}(t)\ =\ 0
\ \ \ \ \ \ \ \ \ \ k\ \ even,\eqno(31)$$
$$\cb{k}{s}{a}(t')\ =\ 0,\ \ \ \ \ \ \ \ \ \ \ \ \cb{k}{s}{a}(t)\ =\ 0
\ \ \ \ \ \ \ \ \ \ k\ \ even,\eqno(32)$$
$$\r{k}{s}{a}(t')\ =\ 0,\ \ \ \ \ \ \ \ \ \ \ \ \r{k}{s}{a}(t)\ =\ 0
\ \ \ \ \ \ \ \ \ \ k\ \ odd,\eqno(33)$$
is a solution of the constraint equations.

In the irreducible case, this theorem reproduces the result of Ref.[3].
But Theorem 2 gives a more precise information about the nature
of the path integral since it relates it
explicitly to the projected kernel of
the operator formalism.
\vskip1cm
{\bf Proof of theorem 1:} The first assertion is obvious: the states
(23) are manifestly annihilated by the minimal BRST charge as well as by
each term of $\Omega^{non\ min}$.  To prove the second assertion, one
checks that the ghost number adds up to zero at each reducibility level.
Consider the level $s$, $0 \le s \le L$, i.e., all the variables
parametrized by
$a_s$.  Assume for definiteness that $s$ is even so that the ghosts
$\eta^{a_s}$ are fermionic.  [If $s$ is odd, then all signs in the following
discussion must be changed.  This does not alter the conclusion
that the total ghost number at level $s$ adds up to zero.]
With $s$ even, $b^k_{s a_s}$ and $\lambda_k^{s a_s}$ (respectively
$\rho_k^{s a_s}$ and $\bar C^k_{s a_s}$) are bosonic (respectively
fermionic) for $k$ even, while they are fermionic (respectively
bosonic) for $k$ odd.  We now count the ghost number:  the state
$|\eta^{a_s} = 0>$ of the minimal sector brings in
$(s+1)m_s/2$.  The state $|b^k_{s a_s} = 0>
|\bar C^k_{s a_s}= 0>$ ($k$ even $\le s$) brings in $[(s-k) + (k-s-1)]\, m_s/2
= - m_s/2$.  The state $|\lambda_k^{s a_s} = 0>|\rho_k^{s a_s}= 0>$
($k$ odd $\le s$) brings in $[(s-k) + (k-s-1)]\, m_s/2
= - m_s/2$.  Hence, the contribution of the non minimal sector,
obtained by summing over $k$ ($0 \le k \le s$), is just $-(s+1)\, m_s/2$.
This compensates the ghost number contribution from the minimal
sector, yielding a ghost number equal to zero for
the states $|q, \; ghosts>$.

\vskip1cm
{\bf Proof of theorem 2:} the proof of theorem 2 proceeds in two steps.
(i) first one verifies than the right and left hand sides of Eq.(26) transform
in the same way under redefinitions of the constraints;

(ii) second, one verifies Eq.(26) in the representation where the constraints
are abelian and the reducibility functions equal to 0 or 1.

The verification of (i) is direct and based on (11-14) above. To check (ii),
one observes that for abelian constraints, $``more"$ in
eq.(8) turns out to be zero, and $\Omega^{min}$ reduces to
$$\Omega^{min}\ =\ \e{A}{0}\ \G{A}\ +\ \sum_{s=0}^{L-1}\ \e{A}{{s+1}}\
\P{\alpha}{s}.\eqno(34)$$
Here, (i) the $G_{A_0}$ are a set of independent (abelian) constraints,
$$\G{a}\ =\ (\G{A}, \G{\alpha}\equiv 0);\eqno(35)$$
and (ii) the reducibility matrices are chosen to be
$$\Z{\alpha}{A}{s}{{s+1}}\ =\ \sim\ \delta^{\alpha_s}_{A_{s+1}},\eqno(36)$$
the other components being zero. We have split the indices $a_s$ as
$$a_s\ =\ (A_s, \alpha_s),\ \ \ \ \ \ \ \ \ \ \ \ A_s=1,...,m_s-r_s,\ \ \ \
\alpha_s=1,...,r_s.\eqno(37)$$
where $r_s = \, rank \; Z^{a_{s-1}}_{a_s}$.
There are no indices $\alpha_L$ since the last
$Z^{a_{L-1}}_{a_L}$ are independent, and there are as many indices $A_{s+1}$
as there are indices $\alpha_s$.

Now, the form of the full BRST charge in the abelian representation
is exactly the same as that of the BRST charge for an {\it irreducible}
gauge system described by the following set of independent constraints:

\noindent
(i) $G_{A_0} = 0$, ${\cal P}_{\alpha_0} = 0$, ...,
${\cal P}_{\alpha_{L-1}} = 0$, with non minimal sector $[b^0_{0A_0}]$,
$[b^0_{1A_1}]$, ..., $[b^0_{LA_L}]$ (we use the
``block notation" $[b]$ of [4] to denote all the variables of a basic
building block $(b, \rho, \lambda, \bar C)$ of the non minimal
sector);

\noindent
(ii) $b^k_{sA_s} = 0$ ($k$ odd) with non minimal sector
$[b^{k-1}_{s-1 \alpha_{s-1}}]$;

\noindent
(iii) $b^k_{s\alpha_s} = 0$ ($k$ odd, $k \, \le \,L-1$)
with non minimal sector
$[b^{k+1}_{s+1 A_{s+1}}]$.

\noindent
The boundary conditions on the ghosts and non minimal
variables are furthermore exactly the same as those that one would
write down by considering the theory as an irreducible
gauge theory with constraints (i)-(iii).

Therefore, if $A_0(q,p)$ is an observable - which we may assume
to commute strongly with the abelian constraints,
$[A_0, G_{a_0}] = 0$ - , the analysis of the
irreducible case applied to the system with constraints (i)-(iii) implies
that
$$<q', \, ghosts|A|q, \, ghosts>\ = \ \ \ \ \ \ \ \ \ \ \ \ \ \ \ \ \ \ \
\ \ \ \ \ \ \ \ \ \ \ \  \ \ \ \ \ \ \ \ \ \ \ \ \ \ \ \ \ \ \ \ \ \
\ \ \ \ \ \ \ \ \ \ \ \ \ \ \ \ \ \ \ \ \ \ \ \ \ \ \ \ \ \ \ \ $$
$$<q', \eta^{\alpha_s} = 0, \lambda^{sa_s}_k = 0|A_0(q,p)\
\prod\ \delta
(\G{A}) \, \delta({\cal P}_{\alpha_i}) \, \delta( b^j_{sa_s})
|q,  \, \eta^{alpha_s} = 0, \lambda^{sa_s}_k = 0>  \eqno(38)$$
up to inessential numerical factors (the index $k$ of $\lambda$ in
(38) is odd) [4, chapter 16].
Since $<u=0| \delta (p_u) |u=0> = (2 \pi )^{-1}$
($u$ even) or $1$ ($u$ odd), one finally gets
$$<q', \, ghosts|A|q, \, ghosts>\ =
\ <q'|A_0(q,p)\ \prod\ \delta
(\G{A}) |q>\eqno(39)$$
up to unwritten irrelevant factors.
The right hand side of Eq.(39) is the projected kernel for the
case of $m$ independent abelian constraints [4]. This achieves the
proof of Theorem 2.

Once Theorem 2 is demonstrated, the proof of Theorem 3 is
immediate : one simply takes for $A$ the evolution operator and
reexpresses $<q', \, ghosts|U(t'-t)|q, \, ghosts>$ as a
path integral.

We have thus established in this letter that the BRST path integral
(27) for the evolution operator, with the boundary conditions
(28)-(33), is a solution of the constraint equations.  We have
actually done more
than merely verifying that (27) solves the constraints (zero is
also a solution of the constraints).  Namely, we
have explicitly related the path integral to the projected
kernel of the evolution operator.  Thus,
the path integral is non trivial and contains
all the information about the evolution of the
gauge system.

\vfill

\eject

\centerline{\bf ACKNOWLEDGMENTS}
M.P.  was supported by  a  grant from the
Universit\'e  Libre  de  Bruxelles.    R.F. and  M.P. wish to thank Professor
R.Balescu and the members of the ``Service de
Physique Statistique, Plasmas et Optique non lin\'eaire"
for their  kind  hospitality.

\vskip4pc
\centerline{\bf FOOTNOTES}
\noindent
$^{\not= 1}$  We stress that in the Dirac quantization
method for constrained systems, {\it all}
the constraints are imposed as conditions on the physical states. There
exists an alternative approach, based in the Fock representation, in which
only the holomorphic part of the constraints is enforced on the physical
states.  This approach is devoid of the scalar product difficulties and
is implemented by means of different boundary conditions on the ghosts [4],
chapter 13.

\noindent
$^{\not= 2}$  It is sometimes stated in the literature
that the equations (3) are inconsistent when the constraints are reducible,
and that the constraints should therefore not be imposed on the physical
states. This is of course incorrect. In the absence of anomalies, the
equations
(3) are consistent both in the irreducible and the reducible cases.  For
example, the constraints for an abelian $2$-form gauge field theory
read $\pi^{ij},_j \ \psi \ = \  0$, i.e. $(\delta \psi / \delta A_{ij}),_j \
= \ 0$.  These equations are not independent and consistent.  Their
general solution is $\psi \ = \ \psi[A^L]$, where $A^L$ is defined through
$A_{ij} \ = \  \epsilon_{ijk} \partial^k A^L \ + \ \partial_i \theta_j
\ - \ \partial_j \theta_i$ (in three spatial dimensions).

\vfill

\eject

{\parindent0pt
{\bf REFERENCES}

1. J.B.Hartle and S.W.Hawking, Phys.Rev. D {\bf 28} (1983) 2960.\par
2. A.O.Barvinsky, Phys.Lett.B {\bf 241} (1990) 210.\par
3. J.J.Halliwell and J.B.Hartle, Phys.Rev. D {\bf 43} (1991) 1170.\par
4. M.Henneaux and C.Teitelboim, {\it Quantization of Gauge Systems,}
Princeton University Press, Princeton, 1992.\par
5. I.A. Batalin and E.S. Fradkin, Ann. Inst. Henri Poincar\'e {\bf 49}
(1988) 145. \par
6. A. O. Barvinsky, Class. Quant. Grav. {\bf 10} (1993) 1895;
A.O. Barvinsky and V. Krykhtin, Class. Quant. Grav. {\bf 10}
(1993) 1957. \par
7. R.Ferraro, M.Henneaux and M.Puchin, J.Math.Phys. {\bf 34} (1993) 2757.\par
8. R. Marnelius,  Nucl. Phys. {\bf B 395} (1993) 647. \par
9. R. Marnelius, Phys. Lett. {\bf B 318} (1993) 92. \par
10. I.A.Batalin and E.S.Fradkin, Phys.Lett. {\bf 122B} (1983) 157.\par
\end